# REPLACEMENT OF MAGNET POWER SUPPLIES, CONTROL AND FIELD-BUS FOR THE PSI CYCLOTRON ACCELERATORS

D. Anicic, T. Blumer, G. Dzieglewski, G. Janser, I. Jirousek, H. Lutz, A.C. Mezger
PSI, CH-5232 Villigen PSI

*Abstract*

Magnet power supplies in the PSI accelerator complex with their control and field-bus are old. Some components are more than 30 years old. To facilitate further maintenance and to meet the more demanding specifications for operation with the 2mA beam, they have to be replaced. The switched power supplies, developed for SLS, will be used. This implies a major redesign of part of the accelerator control system, which is currently based on CAMAC, ROAD-C and other in house developed hardware including the machine protection system. The modified control for the new set-up will be based on VME and alternatively CAMAC, with dedicated processors for the functionality of the machine protection system.

## 1 INTRODUCTION

The everlasting need for renewal of the control system has hit us again. Power-supplies (PS) are old, the field-bus is outdated, the Front-End processors are no longer available and serial CAMAC is not fast enough. Technology of some of these components goes back to the early 1970's. Stability, ripple and mains rejection of the PS is no longer up to the requirements imposed by the 2 mA beam. The new PS's have no longer an analog interface.

All this implies a new control concept for the PS, including the KOMBI, a local controller that also acts as a supervisor to generate Interlock signals, Interlock being our hardware implemented fast run permit system. ROAD-C, the field bus used to control the PS via the KOMBI is old, in house developed and lacks error detection. All I/O in our system is based on CAMAC. We intend to add the possibility of using VME I/O. For some applications the I/O bandwidth of the 5 MHz bit serial loop is a limiting factor.

On the other hand the overall system is very good and flexible to changes. We plan to use the same control system for the PROSCAN project (the new biomedical facility at PSI) under construction. This is motivation enough for a major upgrade of the system.

## 2 MODIFICATION PATH OF SYSTEM

We will extend the control system Ethernet to strategic points in the equipment buildings. 100Mbps optical links with the option to go to 1Gbps and beyond will be used in conjunction with Switches that support diagnostic facilities. The addition of front-ends at these locations adds the possibility of VME I/O in the field. Intelligent interface modules to the PS with a dedicated CPU in VME, will replace the old "KOMBIs" as well as part of the Interlock logic.

This enhancement implies the extension and the port of the front-end software to a new processor.

## 3 THE NEW FRONT END

### 3.1 FEC software upgrade/replacement

In our distributed client-server based Control System, the Front-End Computers (FEC) act as servers, providing the I/O for data acquisition and control. The FECs continuously wait for client requests (Ethernet 802.3, UDP or Unix Message Queues), perform the requested operations and return status and results. FECs are configured at boot time with the data related to Device, Module-handler, Address etc. provided by our configuration database. Module-handlers correspond to different electronic components (modules). Fig.1 shows the FEC software architecture.

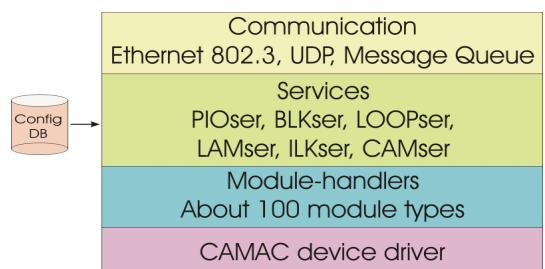

Fig.1: CAMAC based FEC software structure

The present system is entirely CAMAC based with the ROAD-C bus transparently mapped to CAMAC. Currently we support about one hundred different modules. At present, the addition of a new module type requires implementation of a new module handler. In the future we intend to make better use of inheritance with object-oriented methods.

The motivation for the FEC upgrade and replacement program is twofold. Firstly, the present HP rt743 single board VME real-time computers, under the HP-RT operating system (LynxOS), are almost ten years old and will not be supported for a long time anymore. Secondly, we have new projects based on VME. The upgrade path has already been chosen. We will use new VME computers with the LynxOS operating system. We have ordered Motorola MVME 5100 PowerPC boards and the LynxOS 3.1.0 operating system.

The FEC software migration path is set up in a few steps. Initially, the existing software will be ported to the new platform and operating system, opening the door for present FEC replacement as needed. For that purpose we keep the existing VME to CAMAC Serial Highway Driver. This step has already been done and took three weeks for two men. There were no significant problems besides the compiler influenced warning messages and the effort of implementing the CAMAC device driver. We previously redesigned the original HP-RT driver, while porting our SW to Linux. We are porting it back to LynxOS. This software has now to prove itself in long-term operational tests. In a second phase we will add the VME device driver and gradually implement VME Module handlers, as they become needed. This will then provide a mixed CAMAC and VME system.

*3.2 The new FEC database extension*

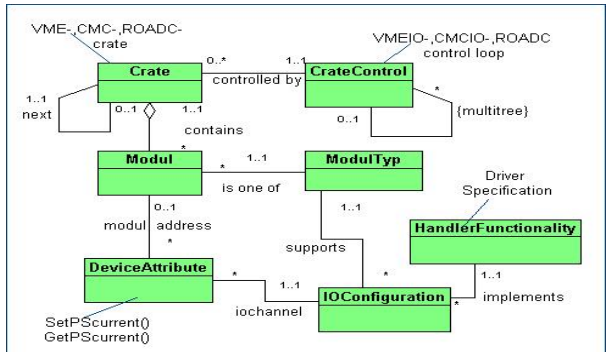

Fig.2: Simplified scheme of I/O-Module configuration

All configuration data used by the Front-End Software is stored in an Oracle Database. The present version only supports CAMAC or ROAD-C based I/O. I/O-Modules are organized by their address: identifier for Front-End, CAMAC Branch, Crate Address in the loop, slot number, and a channel or substation number. Channel numbers are used to identify ADC channels. Substation numbers are used to identify ROAD-C module address lines.

The association or mapping between the I/O and the real device attribute description (e.g. read PS current, set PS current) is normally configured by the Module-handler functionality.

To add the support of VME I/O-Modules both the I/O-Module and the Module-handler mapping part will be extended in the database. New tables, views and their relationships together with the appropriate procedures have to be created, in order to support the new requirements.

For data entry, new application programs are developed using the Oracle Forms Developer tool. The data extraction will use programs written in PRO-C or Java Stored procedures. These programs then supply the Front-End computers with the necessary data to access all connected I/O-Modules. Configuration data contains module identification lists and Device Attribute lists. At Front-End boot time the software first checks for the existence of each I/O-Module and then creates the corresponding device attribute model objects.

## 4 THE NEW PS INTERFACE

The new PS interface is under development in cooperation with industry. HYTEC Electronics will provide a VME IP carrier board (model 8003) with 4 IP-Sites, where they incorporate a Digital Signal Processor (DSP) to allow for local processing.

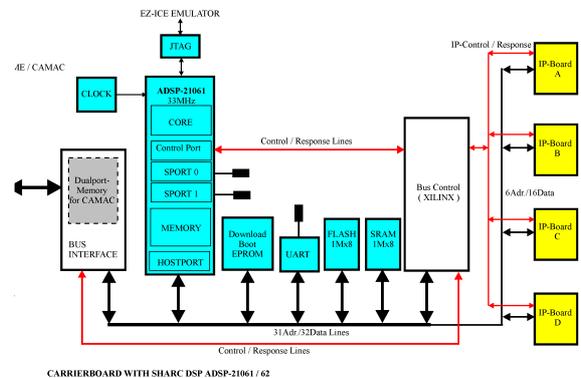

Fig.3: VME-IP Carrier Board

A CAMAC version will also be produced with a dual-port memory acting as interface to CAMAC. The DSP will boot from an internal flash EPROM that can be loaded through VME/CAMAC or by UART (Fig.3). The same Carrier board will also be used for the Interlock logic. Fig.4 shows a structural diagram of PS boards in connection with the Interlock. The DSP provides the data presentation and communication to the Front-end process via memory mapped Registers in its SRAM memory. The on-board processor controls up to six PS, connected to the carrier board, three times two links per IP module are supported. One IP module is used to implement the Interlock functions. The physical connection to the PS is via transition boards and uses fiber-optic cables. The DSP provides the PS supervision to generate Interlocks for current limits, impedance out of tolerance, etc. Up to six PS may be combined to form virtual devices.

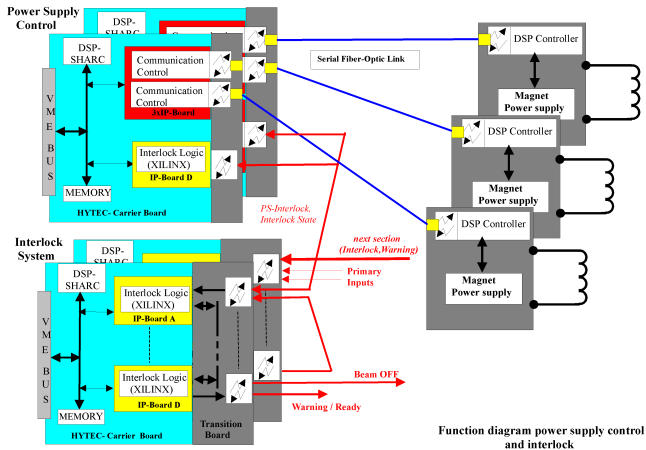

Fig.4: Function Diagram for PS Control and Interlock

Fig.5 shows an overview of the fiber link controller for the point-to-point communication with the PS. On the IP module a XILINX FPGA is used to implement the required logic functions. The opto-electronics part is realized on the VME-Transition board. The FPGA firmware is divided into three logical parts, for two fiber optic links.

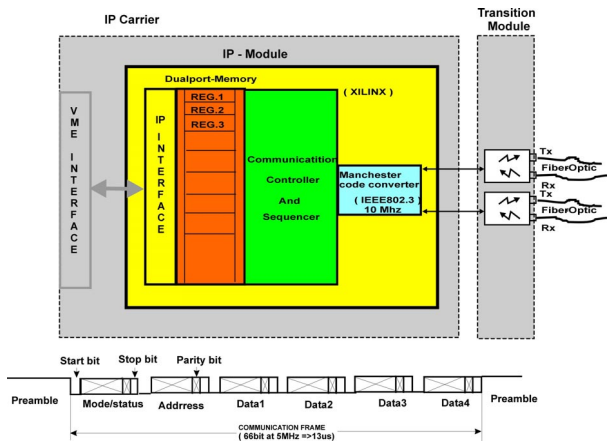

Fig.5: Fiber link hardware and protocol

The protocol of the communication link is always organized in identical blocks (Fig.5). Each block starts with a preamble, followed by a six-byte frame (a byte consist of 1 start bit, 8 data bits, 1 parity- and 1 stop bit). The preamble is a repeated 1 signal for the duration of 12 bits or longer.

The dual port memory used for communication with the FEC is realized in the FPGA. It consists of 64 bytes write-only and of 64 bytes read-only memory for each link. Both the DSP and the VME-CPU may access the memory-mapped registers of the IP-board. A write access will causes the module to send the appropriate data as a command over the optical fiber link to the power supply controller. The answer comes back via the fiber link is stored in the corresponding read register. In addition an interrupt or flag informs the CPU that new data has arrived.

## 5 INTERLOCK

The new Interlock module is composed of the VME carrier board, four IP modules and the transition board shown in Fig.6. One IP module can serve up to ten Interlock channels, four IN, two OUT and six configurable as IN or OUT. All channels are isolated from the XILINX by opto-couplers. The transition board supplies the Interlock channels via DC/DC converters, insuring over-current and over-voltage protection. The new Interlock will also allow resolving pre- and post-trigger of Interlock event conditions, this is required for a refined diagnostic of the causes of an Interlock. In order to allow for the pre- and post-trigger analysis, input channels are equipped with time counters. The same IP Interlock module is also used in the fourth location of the carrier board for PS control.

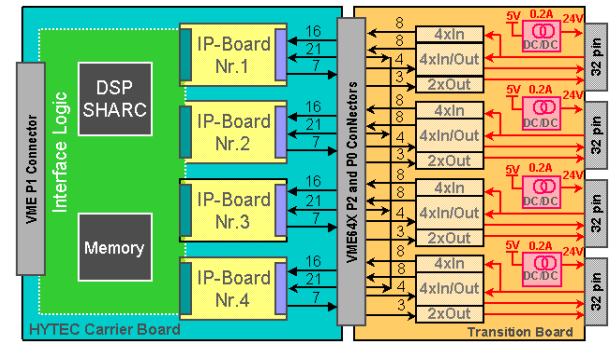

Fig.6: Interlock IP module and transition board

## CONCLUSIONS

We presented an upgrade path to bring our control system up to state of the art again. Both control systems for PROSCAN and for our accelerator will profit from this development.